\documentclass[12pt,preprint]{aastex}
\usepackage{natbib,psfig,aastexug,graphicx}

\begin{document}

\title{Multivariate Monte Carlo Methods for the Reflection Grating
  Spectrometers on XMM-Newton}

\author{J. R. Peterson$^1$, J. G. Jernigan$^2$, S. M. Kahn$^1$}

\affil{$^1$ KIPAC, Stanford University, PO Box 20450, MS 29, Stanford, CA 94309}

\email{jrpeters@slac.stanford.edu, skahn@slac.stanford.edu}

\affil{$^2$ Space Sciences Laboratory, University of California, Berkeley, CA
  94720}

\email{jgj@ssl.berkeley.edu}

\begin{abstract} We propose a novel multivariate Monte Carlo method as an
  efficient and flexible approach to analyzing extended X-ray sources with the
  Reflection Grating Spectrometer (RGS) on XMM Newton.  A multi-dimensional
  interpolation method is used to efficiently calculate the response function
  for the RGS in conjunction with an arbitrary spatially-varying spectral model.
  Several methods of event comparison that effectively compare the
  multivariate RGS data are discussed.  The use of a
  multi-dimensional instrument Monte Carlo also creates many opportunities
  for the use of complex astrophysical Monte Carlo calculations in diffuse
  X-ray spectroscopy.  The methods presented here could be generalized to
  other X-ray instruments as well.
\end{abstract}

\section{The General Problem of Diffuse X-ray Spectroscopy}

Diffuse sources, such as clusters of galaxies, supernova remnants, and the
interstellar hot haloes of elliptical galaxies, comprise some of the most
interesting targets for astrophysical X-ray spectroscopy.  Photons from these sources are
 focused and dispersed by optics and recorded in detectors designed to measure
three interesting quantities:  two related to the position on the detector
($x$, $y$) and one related to an intrinsic photon energy measurement ($p$).
These quantities are indirectly related to the sky coordinates ($\theta,
\phi$) and the energy of the incident photon ($e$) that can be
predicted from an astrophysical model with parameters, \{$\eta_1$, $\eta_2$,
$\eta_3$, \ldots\} $\equiv$ $\eta_i$.

The fundamental goal of data analysis for diffuse X-ray spectroscopy is to
 calculate the probability of a given distribution of photons in the ($x$,
 $y$, $p$) data-space given the parameters of a particular astrophysical
 model.  Prior to the launch of the {\it Chandra} (\citealt{weisskopf}) and
 {\it XMM-Newton} (\citealt{jansen}) observatories, available instruments were
 characterized by sufficiently poor spatial and/or spectral resolution that
 simple one-dimensional spectral fitting techniques applied to data extracted
 from the image were sufficient for most applications.  However, the grating
 experiments and non-dispersive CCD experiments on these two new observatories
 both have significant imaging and spectral capabilities, making it warranted
 to develop new techniques that utilize the full dimensionality of the data.
 In particular, the Reflection Grating Spectrometers (RGS) on XMM-Newton have
 a number of unique characteristics that make them powerful X-ray
 spectrometers for arcminute-size X-ray sources.  In this paper, we demonstrate
 that the complex nature of
 the response function for the RGS requires new analysis methods that utilize
 the full dimensionality of the data.  We also demonstrate how these methods
 might be useful for other X-ray instruments.

The techniques for analyzing an X-ray spectrum of an unresolved source are
well-developed and have been essentially unchanged since the analysis of early
X-ray spectra (\citealt{gorenstein}).  For that case, there is only one
interesting measured quantity that is related to the energy of the photon.  For non-dispersive spectrometers (e.g. CCDs, calorimeters, and proportional counters), the photon energy measurement ($p$) is the useful quantity.  For dispersive spectrometers (e.g. gratings and crystals), the dispersion coordinate, $x$, or the position along the detector parallel to the dispersion axis, is the useful measured value.  In either case, the detection probability, $D$, of finding a photon at a given $x$ or $p$ is then given by
\begin{eqnarray}  
D(x~\mbox{or}~p~\vert~\eta_i ) = 
 \int de~R(x~\mbox{or}~p ~\vert~ e)~S(e~\vert~\eta_i )
\end{eqnarray}
\noindent
which is an integral over all energies, $e$.  Here, R is the response kernel
and S is the input spectral model, which is a function of both the energy and
the input parameters.  The response kernel describes the probability of
obtaining a distribution of measured values given a photon with incident
energy, $e$.  The integral is computed numerically by converting the
expression into a sum of discrete energies in well-developed software
packages, such as XSPEC (\citealt{arnaud}) or SPEX (\citealt{kaastraspex}).

For an extended source, the probability distribution is described by a three-dimensional function and its calculation requires an integration over both the sky coordinates ($\theta,~\phi$) and the intrinsic energy ($e$).  The three-dimensional detection probability, D, is then given by
\begin{equation}
D(x,y,p~\vert~\eta_i ) =
\int de~d\theta~d\phi~R(x,y,p ~\vert~
\theta,\phi,e)~S(\theta,\phi,e~\vert~\eta_i )
\end{equation}
\noindent
Computing this entire integral directly is often impractical.  Nevertheless,
some approximations can be made that are useful in restricted
situations.  In particular, one can assume: 1) the source spectrum is independent
of spatial position, and 2) the response does not vary as function of the
off-axis angle.  The problem can then be reduced to a one-dimensional
integral, as for a point source.  The former assumption is sometimes
justified, but has become untenable in many recent analyses.
For example, modeling an X-ray cluster of galaxies as a thermal plasma whose
temperature varies spatially already violates this assumption.
The second assumption depends on the nature of the X-ray instrument as well as
the angular scale of the problem being investigated.  Attempts to circumvent
this problem by weighting the response matrix for Chandra ACIS-S observations
can be found in \cite{markevitch} and in \cite{arnaud2} for XMM-Newton EPIC
data. We demonstrate in \S2 that the
complex nature of the response function, $R$, for the Reflection Grating
Spectrometer makes this approximation impossible to implement without assuming
that there is no spectral variation over the entire field of view.
Instead, we are forced to consider new methods that consider the full
multi-dimensionality of Equation 2.

Equation 2 can be evaluated directly through a Monte Carlo calculation as we will
  demonstrate in \S3.
  It has been known for some time that Monte Carlo methods are
efficient for the evaluation of multi-dimensional integrals
(\citealt{metropolis}).  Few photons are typically detected in X-ray
  observations, so the exact calculation of the full integral is superfluous.
After constructing an efficient Monte Carlo algorithm, we outline a generic
approach for the analysis of diffuse X-ray sources with Monte Carlo
calculations in \S4.  In \S5, we discuss several methods of event comparison that can
be used with the photons simulated with the RGS Monte Carlo and real data.  In
\S6, we discuss several straightforward extensions to these methods.

\section{Structure of the RGS Response Function}
The Reflection Grating Spectrometers consist of three instrumental
components: the mirror module, the RGS grating array (RGA), and the RGS focal
plane cameras (RFC).  There are two nearly identical sets of each of the three
components.  The full description and calibration of these
components has been covered in \cite{denherder} and is also documented in the {\it XMM Users Handbook}.  We briefly discuss some of
the important characteristics of each of the three components for the purpose of data analysis.

  The mirror module consists of 58 concentrically-aligned gold-coated mirrors
  shells.   Photons hit each Wolter Type I shell twice.  The angular
  resolution is set by surface deformations of the mirrors and the relative
  alignment of the shells.  The X-rays then arrive at
  the reflection grating array (RGA) after exiting the mirror module.
The RGA consists of 182 coaligned rectangular reflection gratings.  Each grating has many triangular replicated grooves spaced at 645.6 lines per
millimeter. 
On-axis X-rays hit the grating at 1.57 degrees and are dispersed
between 2 and 5 degrees depending on their wavelength. The gratings are
precisely arranged in an array, which is aligned so that all light hits
the gratings at the same incidence angle (\citealt{kahn2}).  If $\alpha$
is the angle of incidence of a photon, then the exit angle, $\beta$, is
determined by the dispersion equation:
\begin{equation}
\frac{m \lambda}{d} = \cos{\beta}- \cos{\alpha}
\end{equation}
where $d$ is the grating spacing and $m$ is the integer diffraction order.
The derivation of this relation is straightforward through Fraunhofer
diffraction theory.  The deviation from this ideal relation is not simple,
however, because it depends on the alignment of the array, the surface
imperfections of the gratings, and the achromatic blurring already induced by the mirror module.
The angle of incidence on the grating, $\alpha$, is related to the off-axis angle
relative to the boresight axis, $\theta$, by 
\begin{equation}
\theta=\frac{F}{L} \left( \alpha - \alpha_0 \right)
\end{equation}
where $F$ is the distance between the RGA and the RFC (6.7m) and $L$ is the
focal length of the telescope (7.5m).  For directions perpendicular to the
grooves of the gratings, photons that enter at an angle,
 $\phi$, will exit at the same angle, which we will call cross-dispersion and
designate by the angle, $y$.  The cross-dispersion equation is then simply,
\begin{equation}
y= -\phi
\end{equation}
The photons are detected by the RFC.
Each RFC consists of 9 back-illuminated CCDs placed in a row that detect
individual X-rays.  Each CCD measures 1024 by 384 pixels, which are 27 microns
on a side.  Each CCD consists of two nodes where the charge is clocked
separately for each.  The intrinsic CCD energy resolution, which is set by the
ability to fully sample the energy deposited in the charge cloud, is
sufficient to separate the spectral orders for sets of photons.  We would
otherwise be left with an ambiguity between wavelength and spectral order.
The gain-corrected pulseheight, $p$, is roughly proportional to the
energy,
\begin{equation}
p = e
\end{equation}
\noindent
The CCD array is approximately aligned along the Roland circle in the
RGS design so that the position along the detector array, $x$, is
approximately equal to the exit angle for the grating, $\beta$, after
correcting for the relative locations of the individual CCDs.  For that
reason, we will use $\beta$ and $x$ interchangeably.

The Reflection Grating Spectrometers were designed to have relatively high
dispersion (i.e., large dispersion angles from the focus) in order to
compensate for the fact that the mirrors would blur X-ray sources by 10
arcseconds (\citealt{kahn}).  The soft X-ray spectrum gets dispersed over an
angular range of about 3 degrees.  Thus, in principle if the
spectral resolution is set mostly by the blurring of the mirror,
spectral resolving powers ($\frac{\lambda}{\Delta \lambda}$) near 3 degrees/10
arcseconds $\approx$ 1000 are possible. 
An important aspect of the high dispersion
angle capabilities is that X-ray sources with angular sizes of order the 10
arcsecond blurring also benefit from the high spectral
resolution.  If an X-ray source is larger than the mirror
point-spread-function (PSF), then the resolution is obtained by
differentiating Equation 3 with respect to the off-axis angle,
\begin{eqnarray}
\Delta \lambda  & = & \frac{d}{m}~\sin{\left(\alpha_0 \right)} ~\frac{L}{F} \Delta \phi\nonumber \\
&  \approx & 0.12 \mbox{\AA}~~\Delta \phi \left(\mbox{in arcminutes}\right)
\end{eqnarray}
\noindent
The RGS dispersion therefore has resolving powers 10 times higher than the
energy resolution of a typical CCD in the Fe L wavelength band for a source of arcminute size.

Equation 3, 4, 5, and 6 form the basic {\it first-order} behaviour of the
response function for the RGS and Equation 7 clearly justifies our use of the
RGS with extended sources.  The full response function, however, is far
from this simple.  We will later use Equations 3-6 to interpolate between
elements of the complete response function.  The complete response function is composed of a series of
two and three dimensional functions.  These functions take into account the relative alignment
of the system, the optical scattering properties of the gratings and mirrors, and a
model for the conversion and diffusion of charge in the CCDs.  The response
probability function for the RGS system, $R$, is given by 
\begin{eqnarray}
D(\beta, y, p \vert~  \theta, \phi, \lambda) &= & [ f_{SA}(\beta \vert \phi, \lambda)
  \times  g_{SA}(y \vert \theta, \beta)+ \nonumber \\
 & & f_{LA}(\beta \vert \phi, \lambda) \times g_{LA}(y \vert \theta, \lambda) ] \times
  \nonumber \\
  & &  h(p \vert \lambda, \rm{CCD~node}) \times i( \beta, \phi)
\end{eqnarray}
\noindent
This probability function predicts the distribution of photons with a given
$\beta$, $y$, and $p$ given a model that predicts the wavelength, $\lambda$,
and angular distribution ($\theta$, $\phi$).  Here $\beta$, $y$, and $p$ are
compared with the event values after they have been corrected for the relative
geometry of the CCD locations and the standard gain, offset, and charge
transfer inefficiency corrections.
The six functions, $f_{SA}$, $g_{SA}$, $f_{LA}$, $g_{LA}$, $h$, and $i$, are
all either two or three dimensional and have one output variable for either
one or two input variables.  The reasons for the particular construction of
the RGS response functions is the result of extensive calibration efforts both
before and after the launch of the XMM-Newton observatory.  A full discussion
of the physical theory that is used for the formulation and calibration of the
response function is beyond the scope of this paper (see
e.g. \citealt{rasmussen} or documentation at http://xmm.astro.columbia.edu), but we
will briefly describe the purpose and general properties of the six functions below. 

$f$ and $g$ represent the convolution of the instrument response of the mirror
shells and gratings together.  $f$ encapsulates the dependence of the function
perpendicular to the dispersion direction and $g$ represents the
cross-dispersion dependence.  They are divided in two parts:  the small-angle
(SA) and large-angle (LA) response.  The small-angle functions represent the
unscattered (coherent) X-rays.  Its width in both the dispersion and
cross-dispersion directions is therefore dominated by misalignments of the
gratings and mirror shells as well as correlated errors in the grating's
groove structure.  The large-angle functions represent the scattered light
that exits the gratings at large (degree-scale) angles.  The relative
normalization between the small-angle and large-angle terms is both wavelength
and off-axis angle dependent.  Generally, the small-angle term is two or three
times larger that of the large-angle term. 

The pulseheight function, $h$, is the response of the various CCDs in the
RFC.  A separate response function is calibrated for each CCD as well as for
each node of the CCD.  The exposure map, $i$, is used to keep track of
where the CCDs are located relative to one another as well as to remove
locations of the angular space where there are bad pixels or columns.  Aspect
drift is included in its calculation.  Its
value is usually either close to one or zero depending on whether there is an
active pixel at a given angular position.  In addition to the six functions
there is an overall normalization of the total effective area (units of
$\mbox{cm}^2$) and a total exposure time of a given observation.  The
combination of these normalizations, the six functions, and a model spectrum
in units of photons per $\mbox{cm}^2$ that can vary spatially, can be used
to predict a given number of photons at a value of $\beta$, $y$, and $p$.  We
plot two-dimensional slices of these six functions in Figures 1-6.  These can be
calculated from the Science Analysis System (SAS) task {\bf rgsmcrgen} (a task
we designed explicitly for these Monte Carlo calculations) for the
first five functions and {\bf rgsproc} (the standard analysis pipeline) for the exposure map.

For an observation of an unresolved source, the calculations of the response function is straightforward.  The substitution of equation 8 into
equation 2 results in a integral relation like that of Equation 1.  This is
obtained after integrating over a given pulseheight and cross-dispersion
selection.  This integration only has to be performed once during an
analysis.  This is routinely done in the Science Analysis System (SAS) in the
construction of response matrices in the task {\bf rgsrmfgen} (the standard
response generator).  Then Equation
1 can be used in standard response matrix manipulations in software
packages, such as XSPEC.

For a spatially-resolved X-ray source, complete integration of Equation 2 is
unfeasible.  Furthermore, additional approximations to reduce the
dimensionality of Equation 8 are not possible without assuming that the source
spectrum does not change as a function of spatial position, the response
function does not change significantly at off-axis angles, and the lost events
outside of a given pulseheight and cross-dispersion data selection is not
significantly different than that of a point source, as we have discussed in
\S1.  Although those three approximations can been applied in some global analyses
of extended sources with the RGS (see either \citealt{rasmussen2} (XSPEC model
RGSXSRC) or \citealt{kaastra} (SPEX function)), we wish to avoid these
assumptions in order to allow the spectral model to vary spatially.  This will later give us much more flexibility in astrophysical
modeling.  The solution to the general problem presented here is the direct
Monte Carlo integration of Equation 2 while using a novel scheme for
interpolating between elements of the expression in Equation 8.   We outline this technique as applied to the RGS below.

\section{RGS Monte Carlo Response Calculation}

Monte Carlo codes have always played an important role
in the calibration of X-ray space observatories and in the simulation of
complex observations.  However, they have not achieved widespread use in data
analysis applications, primarily because of slow computation speeds.  Here we
describe a method that generates events at rates of $10^4$ to $10^5$ photons
per second per GHz of processor speed and apply it to the RGS response functions.

A photon can be generated through probability density functions, like the six
functions described in section 2, in the
following way.  Let the probability of a photon being detected with an output
value, $a$, given some input variable, $b$, be represented by a probability
density function, $P(a~\vert~ b)$, normalized to unity.  We want to sample
this distribution by obtaining a set of events with various values of $a$.
First we calculate the cumulative distribution:
$C(a)=\int_{-\infty}^a~da^{\prime}~P(a^{\prime}~\vert~ b)$.  Drawing a random
number, $u$, between 0 and 1 then gives us a photon with a value of $a$ where
$C(a)=u$.  Photons can also be thrown away in order to maintain the proper
normalization if the probability varies as a function of $b$.

We incorporate the above method in a two step process to compute the integral
in Equation 2.  First, photons are chosen using the model function, $S$,
where the input variables, $b$, are the model parameters and the output
variables, $a$, are the photon energy and sky coordinates ($e$, $\theta$,
$\phi$).  We then use a second Monte Carlo to predict sets of detector
coordinates ($x$, $y$, $p$), using the sets of photon energies and sky
coordinates as the input variables, $b$, and involving the response function,
$R$, as the multi-dimensional probability distribution.  The detailed
implementation of the first step depends greatly on the type of astrophysical
model, so we do not attempt to cover all cases here.  The second step involves
only the RGS instrument response, however, and is used in the same way for all
observations and analyses.  In the remainder of this section, we concentrate
on ways of maximizing the efficiency of this second step.

The calculation of a single element of the response kernel, $R$, is
computationally intensive, since $R$ is not described by simple analytic
functions as we have discussed in \S2.  The response kernel can be pre-calculated, however, and stored at
various grid points.   This improves the speed of photon generation by several
orders of magnitude over recalculating the response function for every photon.  The
RGS response function, however, contains many three dimensional functions
requiring billions of elements at full resolution.

\subsection{Interpolation Scheme}

Instead of saving the functions in Equation 8 in fine grids, 
we save the response kernel on coarse grids and
then interpolate between grid points to obtain the intermediate values.
Consider first the one-dimensional case, where we want to know the probability
response function, $P(a~\vert~ b)$, of some variable $a$, at some point, $b$.  Assume that the
response has been pre-calculated at positions $b_1$ and $b_2$ such that
$b_1<b<b_2$.  The response at position $b_1$ can be used as an approximation
for the response at $b$ some fraction of the time and the response function
at $b_2$ can be used the other fraction of the time.  These fractions are
determined by the distance that $b$ is from $b_1$ and $b_2$.   A random
number, $r$, is generated to determine which function to use.  The response at
$b_1$ is used if $r$ is greater than $\left( b-b_1 \right) / \left( b_2-b_1
\right)$.  In this way, the intermediate probability function becomes a linear
combination of the probability function at the two grid points after many
photons are generated.  This procedure is illustrated schematically in the top panel in Figure 7.

An additional step is needed if the response functions have sharp peaks that
shift as a function of $b$.  This is clearly the case for the dispersion
function in Figure 1.  The response peaks, however, can be shifted by
interpolating by the derivative of the response function (i.e., the derivative
of Equations 3-6). Say we choose to use the response at
position $b_1$ by the random number $r$.  Then after choosing a value of $a$
from that response function, we shift $a$ by $a \rightarrow a+\frac{da}{db}
(b-b_1)$.  This shifts the response function as illustrated in the
bottom panel in Figure 7.  These two interpolation steps are, in general, much
faster than the steps needed to calculate the response function.  Such Monte
Carlo interpolation methods are also easily generalized to more dimensions by
drawing several random numbers, $r$, for each dimension and then shifting by
each of the partial derivatives in each dimension.  We will repeat this
interpolation scheme several times using the specific RGS response functions in the following section. 

\subsection{Detailed Response Calculation Procedure}

There are several steps in the Monte Carlo calculation that involve
manipulating the functions described in \S2 by the method outlined in \S3.2.  The goal of this
calculation is to start with a set of sky angles ($\theta$, $\phi$) and
wavelength, $\lambda$, for a given photon and end with a set of detected event
coordinates, ($\beta$, $y$, and $p$).
The first step of the RGS Monte Carlo is to choose one of the two
instruments.  This is accomplished by taking the sum of the normalization for $f_{SA}$ and
$f_{LA}$ for RGS 1 and 2 separately.  Then we draw a random number between 0
and 1 and choose the instrument based on the relative values of the two
normalizations. 

  The second step is the correction for the instrument boresight.  The RGS
instruments are misaligned with respect to one another, so the sky
coordinates, $\theta$ and $\phi$, are not trivially related to the $\theta$ and $\phi$ that we use in
the instrument response functions.  We therefore define $\theta_{ST}$, $\phi_{ST}$ as the
sky angles relative to the star tracker and then convert these coordinates to
a set of coordinates ($\theta$, $\phi$) for each RGS instrument.  This is
accomplished by a standard Euler transformation, such that the rotated
coordinates, $r^{\prime}$, are related to the input vector, $r$, by

\begin{equation}
  r^{\prime}_i = M_{ij} r_j
\end{equation}

\noindent
where $M$ is an approximately diagonal rotation matrix specified for each RGS.
More details of this transformation can be found in \cite{peterson3}.

A third step chooses whether the photon follows the small-angle or large-angle
response distribution.  This is simply accomplished by taking the relative
normalization of the $f_{SA}$ and $f_{LA}$ distribution and choosing based on
that distribution.  Assume first that we have chosen the large-angle
distribution.  Then, we find a $\beta$ for that photon by the following
procedure.  First, find the closest wavelength on the wavelength grid that
defines $f_{SA}$.  Define the difference between the closest wavelength,
$\lambda_1$, and the
desired input wavelength, $\lambda_0$, as $\Delta \lambda_{10}$ and the difference
between the second closest wavelength, $\lambda_2$, and $\lambda_0$ as $\Delta
\lambda_{20}$.  Then we will use the reference wavelength, $\lambda_1$, if a
random number is less than $\frac{\Delta \lambda_{20}}{\Delta \lambda_{10} +
  \Delta \lambda_{20}}$.  Otherwise, we will use the reference wavelength
$\lambda_2$.  Define the reference wavelength that we have chosen as
$\lambda_i$. 

We repeat the same procedure for finding a reference $\theta$, which we define as $\theta_i$.  Then we look up the cumulative probability
distribution, $f_{SA}$, which is a function of $\beta$, for a given $\theta_i$ and
$\lambda_i$.  It will have a monotonically increasing value between 0 and 1 at
each $\beta$.  Note that the maximum value might be less than 1 because at a given
$\lambda$ and $\theta$ since its value might be lower than at another $\lambda$ and
$\theta$ where the response is higher.  Then another random number, $r$, is chosen and
we find the value of $\beta$ where $r=f_{SA} (\beta~\vline~\lambda_i,
\theta_i)$.  If the value of $r$ is greater than the maximum value for $f_{SA}$
then the photon is thrown away and we start the procedure over again.  We,
however, keep track of the number of photons that have been discarded.

Assume now that we have successfully chosen a value of $\beta$=$\beta_0$.  Then,
the final value of $\beta$ that will be output by the Monte Carlo is defined
by 
\begin{equation}
\beta=\beta_0-\frac{\cos{(\beta_0)}-\cos{(\alpha_0+\theta_0\frac{L}{F})}}
{\sin{\beta_0}}\frac{\lambda-\lambda_0}{\lambda_0}+\frac{\sin{(\alpha_0+\theta_0\frac{L}{F})}}{\sin{(\beta_0)}}(\theta-\theta_0)\frac{L}{F}
\end{equation}
\noindent
This equation is derived by differentiating the dispersion equation.
A final step that avoids some aliasing is to shift that value of $\beta$ on to
a uniform $\beta$ grid according to the same procedure where we found the
reference $\theta$ and $\lambda$. 

The fourth step consists of choosing the cross-dispersion value, $y$, based on
the function, $g_{SA}$.  We first choose the reference $\phi_i$ based on the
same procedure we used to get the reference $\lambda$ or $\theta$ in the
previous step.  We also use the value of $\beta$ we obtained from the previous
step.  Then a random number, $r$, is chosen between 0 and 1 where we will find
a value $y$ such that $r=g_{SA} (y~\vline~ \phi_i ,\beta)$.  There is some
possibility that we will throw out this photon for sufficiently high values of
$r$.  Otherwise, we will end up with a value $y_0$.  This value is shifted
according to the equation,
\begin{equation}
y=y_0+\left(\phi-\phi_i\right)
\end{equation}
\noindent
Finally, we align $y$ according to a predefined reference grid according to
the same procedure we have used to find our reference $\theta_i$.  We now have
a prediction for both the dispersion variable, $\beta$, and the
cross-dispersion variable, $y$.

If we had chosen the large-angle distribution instead of the small-angle
distribution at the beginning of step 3 the procedure is nearly identical.  We
first use the distribution $f_{LA}$ instead of $f_{SA}$ to get the value of
$\beta$.  Then we use $g_{LA}$ instead $g_{SA}$ to get the value of $y$.
$f_{LA}$ depends on wavelength, however, rather than $\beta$.  We then have a
prediction for the values of $\beta$ and $y$ for the photons that are produced
from the large angle part of the response as well.

The fifth step consists of using the exposure map to determine if the
particular value of $\beta$ and $y$ corresponds with an active region of one
of the CCDs.  Each
value of the exposure map, $e$ will have a value between 0 and 1.  We then simply
find the value of $i(\beta, y)$ and throw the photon away if the value of the
exposure is less
than a given random number chosen between 0 and 1.  Otherwise, the photon is
considered detected and we proceed to the final step.

The last step consists of choosing the CCD pulseheight distribution.  The
distribution is different for each of the two nodes for each CCD.  Therefore
using the chosen value of $\beta$ we determine which CCD node for the photon
 using a simple linear function.  This is correct in the limit that
the CCDs are approximately aligned in $\beta$, which is a good approximation
although they are slightly rotated.  Then, we choose a reference wavelength, $\lambda_i$,
based on the wavelength grid for the function $h$ using the same procedure used in step 3 and 4.  A
random number, $r$, is chosen between 0 and 1.  The predicted CCD pulseheight
is then chosen by finding the value of the pulseheight distribution such that
$r=h(p~\vline~\lambda_i, \mbox{CCD~node})$.  We then shift the value of the
pulseheight by 
\begin{equation}
p = p_0 - \frac{hc}{\lambda} +\frac{hc}{\lambda_i}
\end{equation}
\noindent
 If $r$ is greater than the value
of $h$ then we throw the photon away.  Otherwise, we have successfully
predicted a value of $\beta$, $y$, and $p$.

The overall normalization is achieved by using the total number of photons
we attempted to simulate, $N$, the maximum value of the effective area, $A$, the
maximum value of the exposure map, $T$, and the relative normalization, $Q$, of the model in units
of photons $\mbox{cm}^{-2}~\mbox{s}^{-1}$.  If we simulate $m$ photons and
there are $n$ photons in the data set, then the normalization of the spectral
model is given by
\begin{equation}
\mbox{Normalization} = \frac{N}{Q~A~T} \times \frac{n}{m}.
\end{equation}
\noindent
The Monte Carlo generates photons at rates of $5000$ photons per second per
GHz of processor speed.  An example of a simulation of monochromatic light
from a point source is shown in Figure 8 and 9.  A public version of this code
is currently being prepared.

\section{Towards a General Method for Diffuse X-ray Spectroscopy}

Given an efficient instrument response Monte Carlo, how can it be used with
the measured data to constrain an astrophysical model?  Schematically, the
following steps are required:

\begin{itemize}

\item{Models are formulated in terms of a set of parameters.  These parameters
  predict probability distributions for the spectral and spatial distributions.}

\item{Photons are then drawn from these probability distributions and assigned
  an energy and two sky coordinates ($\theta$, $\phi$).}

\item{The detector coordinate and pulseheight values ($x$, $y$, $p$) are predicted from the instrument Monte Carlo. }

\item{The simulated events are compared with the measured photon events via a
  comparison statistic.}

\item{  Finally, an iteration is performed to optimize this
  statistic subject to variation of the input parameters.  When the iteration
  converges, the best fit has been found.}

\end{itemize}

\noindent
These steps are also outlined in Figure 10 and are similar to standard
analysis methods practiced in high energy astrophysics.
An example of a simulation and an actual data set for a galaxy cluster is in Figure 11 and 12.

\section{RGS Event Comparison}

Following, we discuss several useful methods of comparing the raw RGS data and
the simulated photons from the Monte Carlo calculation like that in Figure 11
and 12.  The obvious advantage
of using a Monte Carlo is that we can manipulate and select the data in any
way since we can perform the same operations to the simulated photons.  For
this reason there are a variety of approaches to event comparison.  We have
found that some are more useful depending on the situation.

\subsection{Wavelength, Order, and Cross-dispersion Assignment}

After simulating a set of photons with a set of values of $\beta$, $y$, and
$p$, it is more convenient to convert those coordinates into a corrected
wavelength, $\lambda^{\prime}$, cross-dispersion value, $\phi^{\prime}$, and
spectral order (the $\prime$ designates that this may not be the true wavelength
but the one measured by the RGS). This is accomplished by using Equation 3 with the value of $\beta$ and cross-dispersion relation with the
value of $y$ after using the pulseheight $p$ to determine whether the photon
falls into first, second, or third order.  We assume a nominal dispersion
coordinate, $\theta$, for all of the photons even though for an extended source 
we may not have the same input $\theta$ for every event.
We can do this because we can perform the same operations the simulated data.
We can interpret a shift in wavelength from a spectral line as either an actual
wavelength shift or a shift caused by the finite extent of the source.

\subsection{Data Selection:  Order selection and cross-dispersion cuts}

A simple yet important step of a Monte Carlo is the selection of data.  We can
perform the same data selection cuts that are normally performed on point
source observations.  A selection in cross-dispersion is achieved by requiring all values of
$\phi^{\prime}$ between an arbitrary $\phi_1$ and $\phi_2$ such that $\phi_1 <
\phi^{\prime} < \phi_2$.  Similarly, a joint pulseheight-dispersion cut is
achieved by requiring that the pulseheight, $p$, falls within the window
$\frac{hc}{\lambda^{\prime}} \pm \sqrt{a+b \frac{hc}{\lambda^{\prime}}}$ for
some arbitrary constants $a$ and $b$.  With extended sources, the joint
dispersion-pulseheight cut is also broadened by calculating the above formula
based on two values of $\theta$ and then making sure the pulseheight value in
within in the window for both values of $\theta$. 
  These selection cuts are important to reduce the number of background events.  The advantage of using the
Monte Carlo is that events can quickly be sorted and removed if they do not
satisfy the selection criteria.  The selection can be changed arbitrarily
without re-running the simulation.  Cross-dispersion cuts, in particular, can be
used to compare different spatial regions.

\subsection{Two-Sample $\chi^2$}

Various statistics can be constructed to determine the quality of the model
used in the simulation.  In particular, a useful specific form of the $\chi^2$ statistic for two samples of data can be computed.  This is given below for binned data where the number of events in the $j$th bin for the data photons is $V_{1j}$ and for the model photons is $V_{2j}$.  If there are $n$ total events in the data and $m$ simulated events in the model then the $\chi^2$ statistic is given by
\begin{equation}
\chi^2=\sum_j \frac{~\vert~ V_{1j}-\frac{n}{m} V_{2j} ~\vert~ ^2}{\frac{n}{m}(V_{1j}+V_{2j})} .
\end{equation}
\noindent
Note that the two sample $\chi^2$ value approaches the familiar continuous form of
$\chi^2$ when $V_{2j}$ is replaced with $m P$ where P is the probability
model and $m \rightarrow \infty$.

The two-sample $\chi^2$ is used to compare both the extracted spectra (a
binned histogram of $\lambda^{\prime}$) and the cross-dispersion distribution
(a binned histogram of $\phi^{\prime}$).  They are both extremely useful in
comparing the spectrum to the model spectrum and the cross-dispersion
distribution compared to the predicted one.  In \cite{peterson} and \cite{peterson2} this method
was employed to compare several spectra and the cross-dispersion distribution iteratively.
It is also possible to compute a two-dimensional $\chi^2$ statistic on the
binned wavelength-cross-dispersion data space.  This is only possible with
extremely bright sources, however, since there are usually a few counts per bin.

\subsection{Two-Sample Cram\'{e}r von Mises}

We have also found an alternative statistic useful when using comparing RGS data.   The
Cram\'{e}r-von Mises statistic, $W^2$ (\citealt{Anderson}), which is a more robust version of the Kolmogorov-Smirnov statistic (\citealt{Smirnov}), is determined by computing the cumulative distribution of the data values, $C_1$, and the model values, $C_2$, and then evaluating the
following sum at each of the model and data values,
\begin{equation}
 W^2 = \frac{m n}{{\left( m + n \right)}^2}
\sum_{{\rm{data~and~model~values}}} {\vert C_1 - C_2 \vert}^2 .
\end{equation}
\noindent
The statistic is most easily computed by sorting the data and model values.
The statistic can be extended to multi-dimensional distributions by comparing
values that are a linear combination of the value in each dimension.  If we
compute the Cram\'{e}r-von Mises statistic of $v$ where $v= a_1 \beta+ a_2 y +
a_3 p$, then we can compare the multi-dimensional distribution.  The
Cram\'{e}r-von Mises statistic is computed for a set of several $(a_1, a_2,
a_3)$ so that the value in one-dimension is emphasized more than other
dimensions in each computation.  If enough combinations are used, the
statistic is relatively insensitive to our choice of $a_i$.  We have used this method to compare the
multi-dimensional distribution of the background.  It works well for this
purpose since few photons fill each bin of the three-dimensional ($\beta$,
$y$, $p$) space.  It may be possible to construct other useful two-sample
multi-dimensional statistics as well and clearly there is more progress to be
made in this area.  

\section{Extensions of the Method}

Below we discuss several extensions to the method.

\subsection{Iteration Scheme}

We have not yet discussed methods for iterating the model parameters after
the photons have been simulated and the statistics have been calculated.  One
method involves changing the parameters of the model after all the photons
have been simulated.  The standard techniques for the iteration of a set of
model parameters without using a Monte Carlo are applicable here as well.
Simplex (\citealt{O'Neill}), simulated annealing, Markov chain Monte Carlo
techniques, or robust grid searches may be the best methods since they all
avoid using derivatives of the fitting statistic and therefore avoid the statistical
$\sqrt{m}$ fluctuations of the Monte Carlo.
	
A more advanced form of iteration involves selecting individual photons
produced from a set of parameters that improve the statistic.  The statistic
is evaluated after every photon is produced instead of after the entire set is
simulated.  This is then the Monte Carlo analog to deconvolution as opposed to
model fitting.  Further discussion of this technique is described in \citet{Jernigan1} and \citet{Jernigan2}.

\subsection{Error Analysis}

Systematic errors dominate over statistical errors in many of the
multi-dimensional global fits we have considered here.  Additionally, the
statistics we have discussed in \S5.1 usually have a
non-universal distribution.  Our focus in using these statistics is merely for
model iteration and not for hypothesis testing or for constructing parameter
confidence regions.

Standard parameter estimation techniques, however, can still be
applied when dealing with the one-dimensional spectrum or a specific feature
in the data.  For example,
one can first calculate $\chi^2$ on the one-dimensional extracted spectrum of both the
data and simulation as in \S4.2.  Then the
difference in $\chi^2$ when a parameter is varied can be estimated.  This
method is
identical to standard parameter estimation techniques in X-ray
astronomy (\citealt{lampton}).  A simple example of this is shown in Figure 13
and 14.  In some other
circumstances, however, bootstrapping techniques
(\citealt{Efron}) could be used to construct the distribution of these multivariate statistics when statistical errors dominate.

\subsection{Other Instruments and Background}

Although we have focused on a number of multivariate Monte Carlo methods for
the Reflection Grating Spectrometer, there is little in our discussion that
could not be generalized to analysis of data from other instruments as well.
The three dimensional structure of the response calculation is not unlike the
structure of a non-dispersive instrument response that has an energy-dependent
point-spread function, vignetting function, CCD response function, and an
exposure map.  The flexibility in astrophysical modeling of this approach may outweigh the advantages of the simplicity of traditional
one-dimensional spectral extraction techniques at least in some situations.  We have also applied this
method to a Monte Carlo of the RGS background induced by charged particles and other
instrumental sources.  The background Monte Carlo has its own response
function that differs from equation 8.  For more details about this Monte
Carlo see \cite{peterson3}.

\subsection{Further Astrophysical Modeling}

After the instrument response is formulated as a Monte Carlo it is reasonable to
consider more complicated astrophysical modeling that could involve a Monte
Carlo approach.  A flexible method is to formulate the astrophysical models
always in three dimensions and then project the photon velocity shifts and spatial positions to the two dimensional sky coordinates.  Radiative transfer also becomes a
straight-forward problem since individual photons can have their frequency
shifted or trajectory altered.  Monte Carlo techniques for this are
well-developed, but these techniques can be used naturally in conjunction
with a Monte Carlo treatment of the instrument model as in \cite{xu}.  We expect that many of the methods discussed here could allow a closer connection between future observational and theoretical work.

\subsection{Acknowledgements}

We thank an anonymous referee for many important comments that greatly
improved this manuscript.  JRP acknowledges many helpful discussions and
technical improvements to the Monte Carlo code and methods by Jean Cottam,
Maurice Leutenegger, Frits Paerels, Andy Rasmussen, Doug Reynolds, Masao Sako,
Cor de Vries, and Haiguang Xu, as well as additional support from the XMM-RGS
team and the XMM collaboration.  SMK acknowledges partial support for
this work under SAO Chandra grant GO2-3178X.

\begin{figure}
\begin{center}
\includegraphics[width=3in]{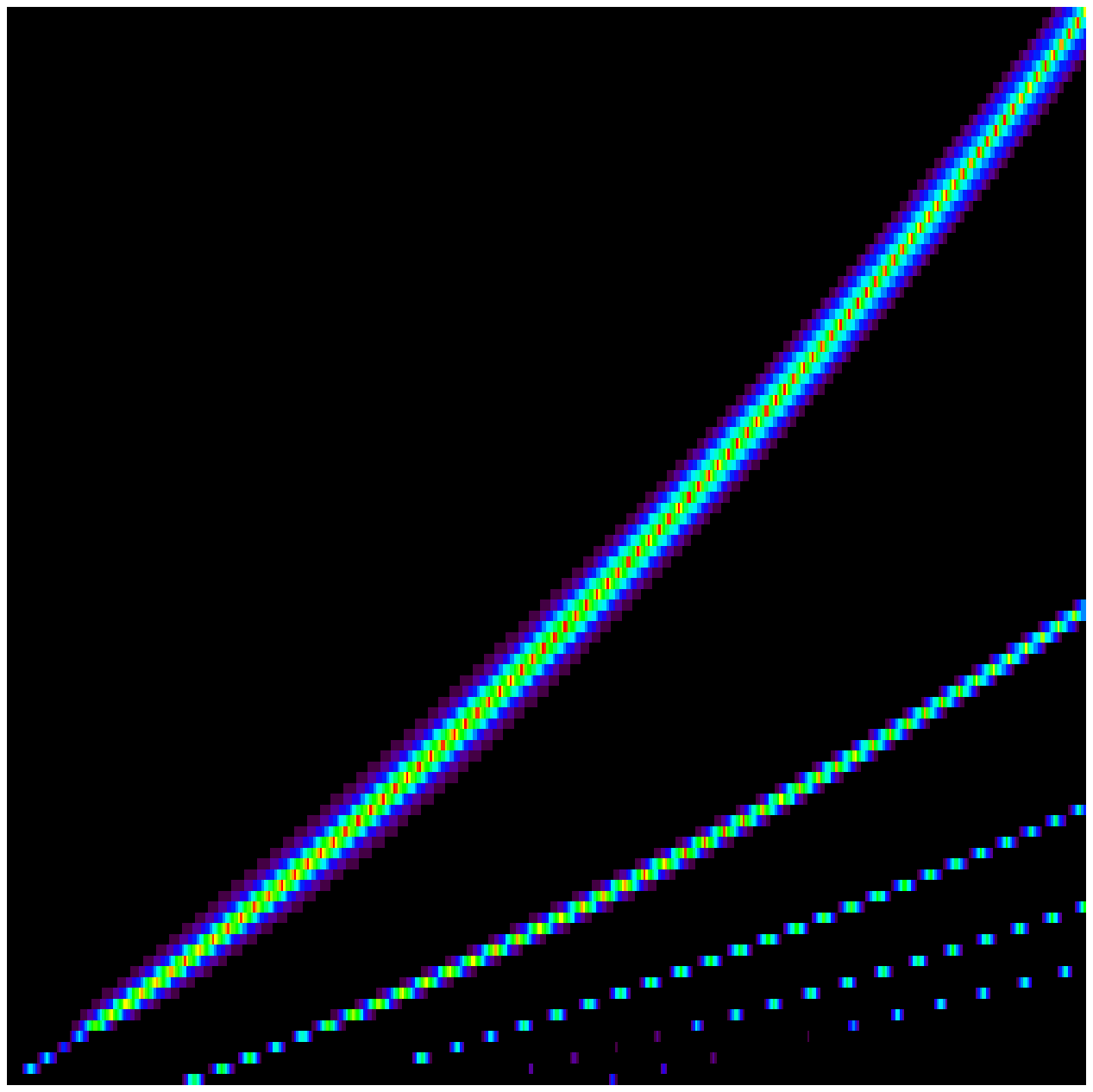}
\end{center}
\caption[Small-angle dispersion distribution]{Response for the small-angle
  part of the RGS dispersion response.  The plot shows input wavelengths
  (vertical axis) and $\beta$ (horizontal axis).  Each line corresponds to a
  spectral order (first through fifth), the brightest one being the first
  order response.  This plot is only the on-axis response.  It varies as a
  function of off-axis position.  The red region is where the response
  function is the
  highest, whereas the black regions are where the response function is low.}
\end{figure}

\begin{figure}
\begin{center}
\includegraphics[width=3in]{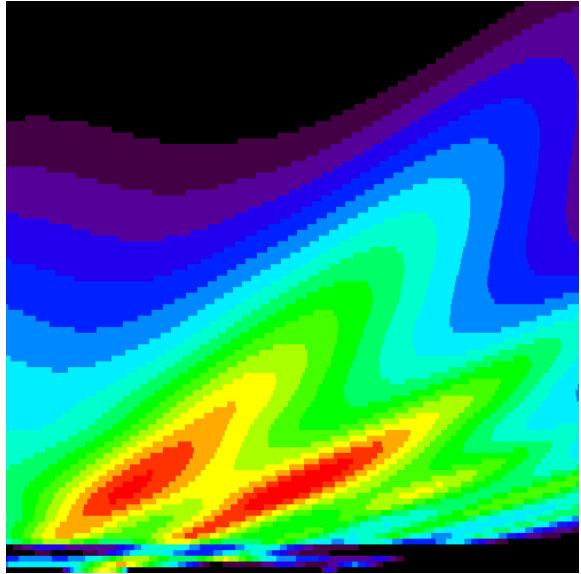}
\end{center}
\caption[Large-angle dispersion distribution]{Response for the large-angle
  part of the RGS dispersion response.  The plot shows input wavelengths
  (vertical axis with wavelength increasing) and $\beta$ (horizontal axis).
  Each peak corresponds to a
  spectral order (first through fifth), the brightest one being the first
  order response.  Contrast this with the previous plot.  This plot is only
  the on-axis response.  It varies as a function of off-axis angle.  The red region is where the response
  function is the
  highest, whereas the black regions are where the response function is low.
  The complex nature of the response function at the bottom of the plot is due
  to the gold M shell edge, which affects the reflectivity dramatically.}
\end{figure}

\begin{figure}
\begin{center}
\includegraphics[width=3in]{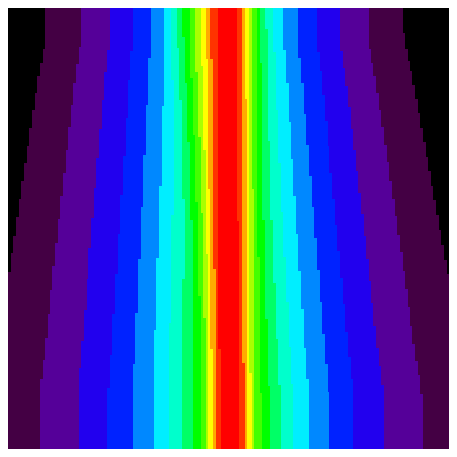}
\end{center}
\caption[Small-angle cross-dispersion distribution]{Response for the small-angle
  part of the RGS cross-dispersion response.  The plot shows input $\beta$
  (vertical axis with $\beta$ increasing) and cross-dispersion, $y$ (horizontal axis).  This plot is only
  the on-axis response.  It varies as a function of off-axis angle.  The red region is where the response
  function is the
  highest, whereas the black regions are where the response function is low.}
\end{figure}

\begin{figure}
\begin{center}
\includegraphics[width=3in]{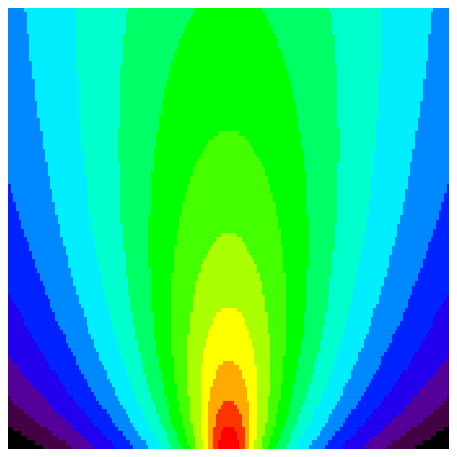}
\end{center}
\caption[Large-angle cross-dispersion distribution]{Response for the large-angle
  part of the RGS cross-dispersion response.  The plot shows input wavelength
  (vertical axis with wavelength increasing) and cross-dispersion, $y$ (horizontal axis).  This plot is only
  the on-axis response.  It varies as a function of off-axis angle.  The red region is where the response
  function is the
  highest, whereas the black regions are where the response function is low.}
\end{figure}

\begin{figure}
\begin{center}
\includegraphics[width=3in]{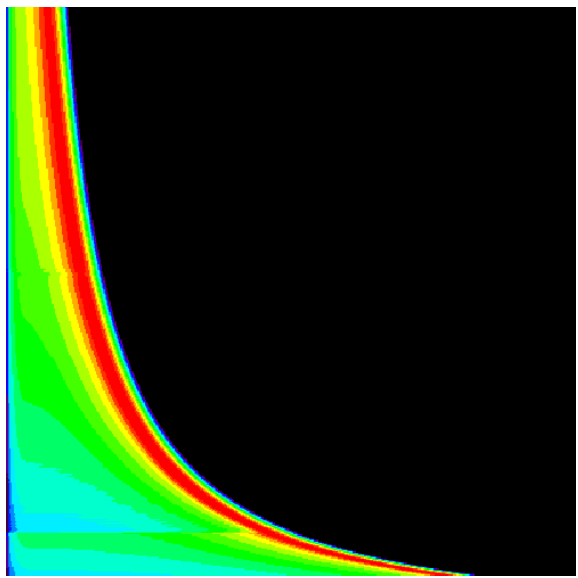}
\end{center}
\caption[CCD pulseheight distribution]{Response for the CCD pulseheight
  redistribution for a particular CCD node.  The plot shows input wavelength
  (vertical axis with wavelength increasing) and pulseheight (horizontal
  axis).  The red region is where the response
  function is the
  highest, whereas the black regions are where the response function is low.
  The horizontal shift in the response near the bottom of the plot corresponds
  with the silicon K shell edge. }\end{figure}

\begin{figure}
\begin{center}
\includegraphics[width=3in]{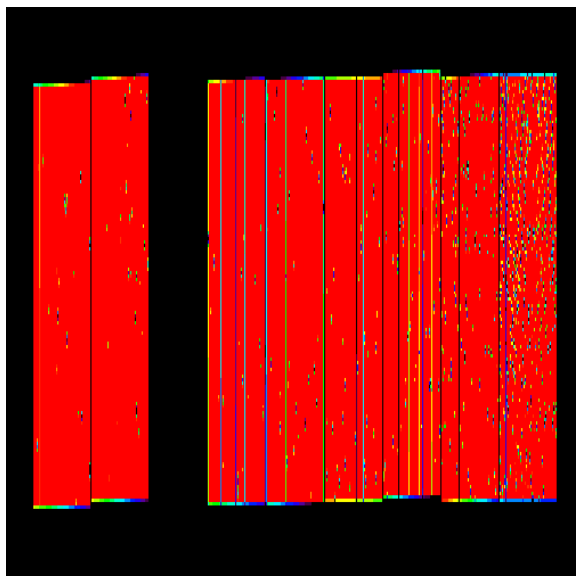}
\end{center}
\caption[Exposure Map]{CCD exposure map showing the positions of the 9 CCDs.
  Unexposed areas included areas where there are hot pixels (points), hot
  columns(lines), and a failed CCD (missing area three from the left).  The
  exposure map does not have sharp edges due to the correction of the small
  aspect drifts of the telescope.  The exposure map is close to 1 in the red
  regions, whereas the exposure map is close to 0 in the black regions.}
\end{figure}

\begin{figure}
\begin{center}
\includegraphics[width=6in]{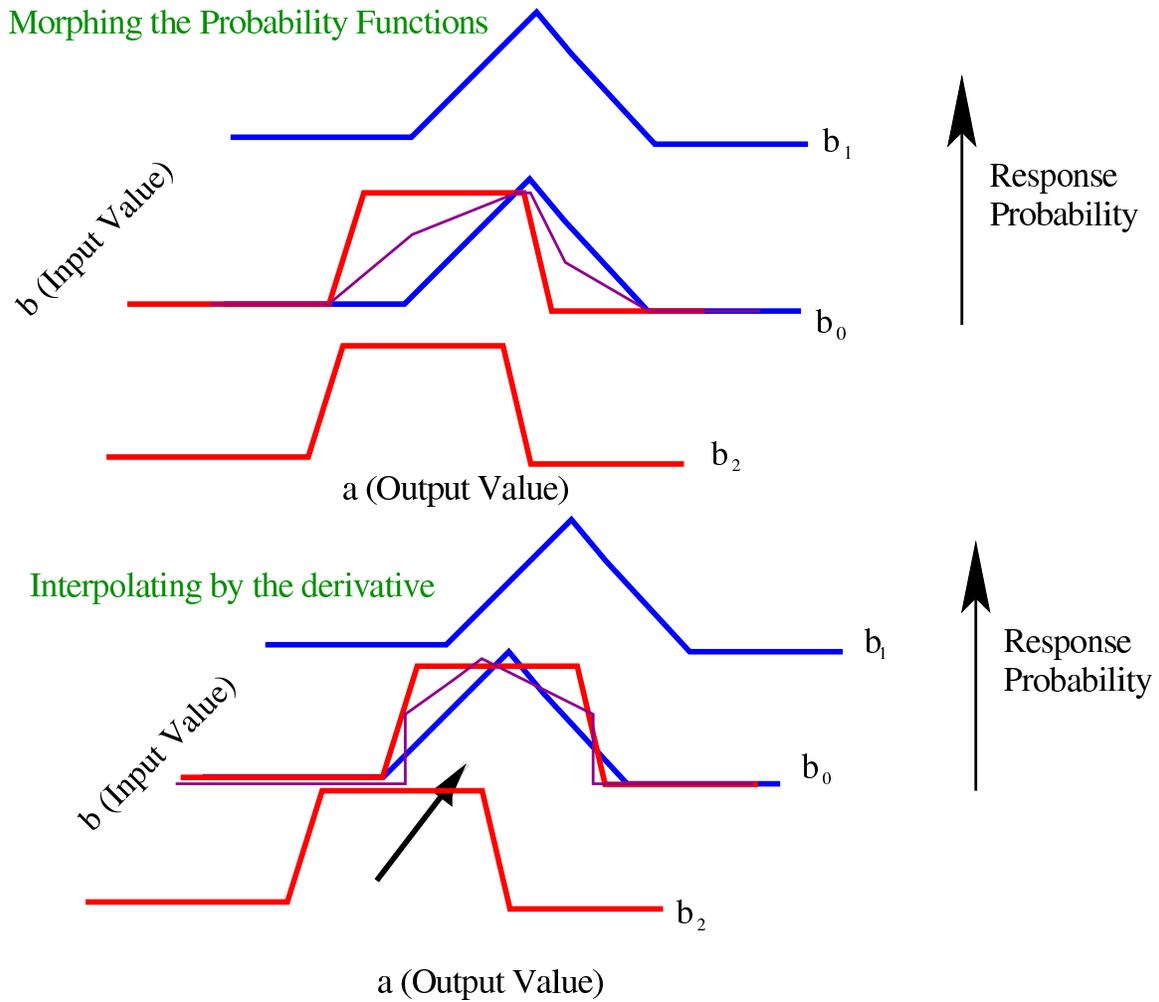}
\end{center}
\caption[Monte Carlo interpolation method]{Schematic of the Monte Carlo
  interpolation method outlined in the text.  In each of the two panels, we
  show a hypothetical response probability for a variable, a, that is
  dependent on some input variable, b.  Instead of calculating the response
  probability 
  for all possible values of b, we assume we have only calculated the response
  probability 
  at $b_1$ and $b_2$ and then use an interpolation scheme to approximate the
  response probability at $b_0$.  If the response peaks line up, we can simply
use the probability distribution at $b_1$ and $b_2$ some fraction of the
time as shown in the top panel and get a combined probability curve shown by
the purple curve.  The probability distributions can also be shifted if the
response probability peaks do not have a common value as shown in the second panel. }
\end{figure}

\begin{figure}
\begin{center}
\includegraphics[width=3in,angle=270]{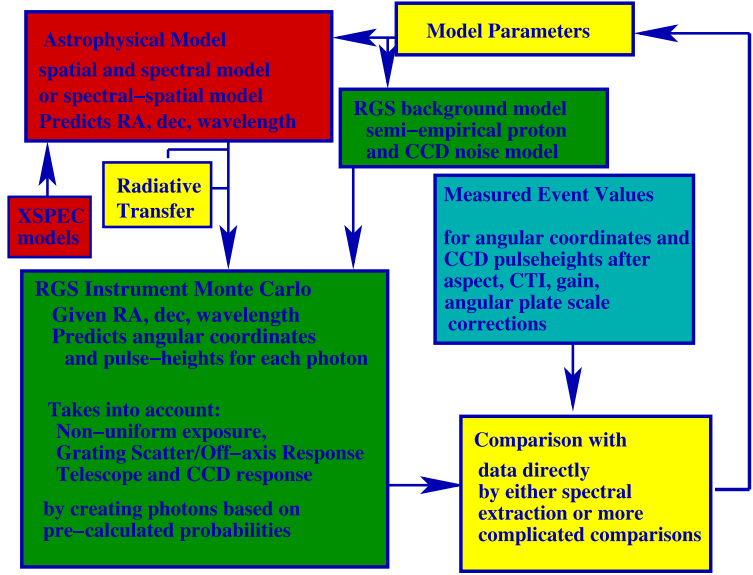}
\end{center}
\caption[Flowchart of the Monte Carlo method]{Flowchart of the Monte Carlo
  method used in this paper.  See section 4 for a detailed explanation of
  each step.}
\end{figure}

\begin{figure}
\begin{center}
\includegraphics[width=3in]{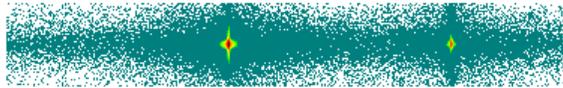}
\end{center}
\caption[RGS monochromatic on-axis response image plot]{Simulation
  of a set of 10 $\mbox{\AA}$ photons.  The plot shows the cross-dispersion
  vs. dispersion histogram.  The left peak is the first order spot and
  the right peak is the second order spot.  Large scattering wings are present
  off a sharply peaked core.  Red shows the high intensity regions, and green
  shows the low intensity regions.}
\end{figure}

\begin{figure}
\begin{center}
\includegraphics[width=3in]{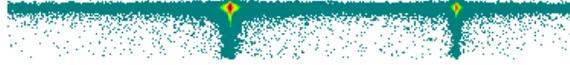}
\caption[RGS monochromatic on-axis response beta-pulseheight plot]{Simulation
  of a set of 10 $\mbox{\AA}$ photons.  The plot shows the pulseheight
  vs. dispersion histogram.  The left peak is the first order spot and
  the right peak is the second order spot.  Large scattering wings are present
  off a sharply peaked core along the dispersion direction and the partial
  event tails of the pulseheight distribution extend vertically.  Red
  represents the high intensity regions, and green represents the low
  intensity regions.}
\end{center}
\end{figure}

\begin{figure}
\begin{center}
\includegraphics[width=3in]{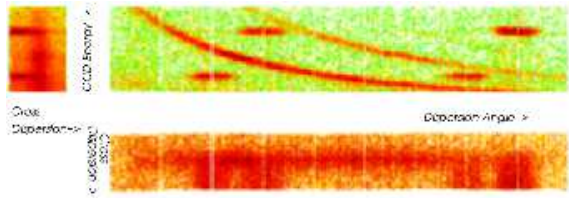}
\end{center}
\caption[Raw RGS data for S\'{e}rsic 159-03]{ Raw RGS data for the galaxy
  cluster, S\'{e}rsic 159-03.  The plot consists of three panels for each of
  the two-dimensional projections of the three dimensional data.  The
  dispersion coordinate vs. cross-dispersion coordinate shows the dispersed
  spectral image.  It is blurred in the cross-dispersion direction due to the
  size of the source.  The three curved lines in the dispersion coordinate
  vs. CCD energy plot show the first, second, and third order dispersed
  spectra.  The four horizontal lines are the Al K and F K calibration
  sources.  Most of the photons in this image are due to Bremsstrahlung.   A
  darker region in the first order curved line is due to Fe L lines.  The red
  regions are high intensity regions, and the green regions is low intensity regions.}
\end{figure}

\begin{figure}
\begin{center}
\includegraphics[width=3in]{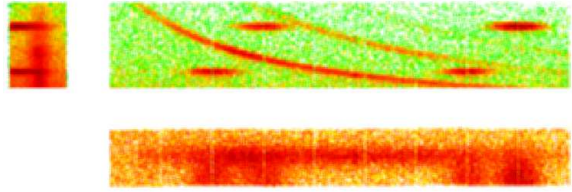}
\end{center}
\caption[Model of RGS data for S\'{e}rsic 159-03]{The same plot as Figure 11
  except the photons are simulated by a model using a Monte Carlo.  Details of
  the simulation can be found in Peterson et al. (2003a).  The green represents
  low intensity regions, and the red represents high intensity regions.}
\end{figure}

\begin{figure}
\plottwo{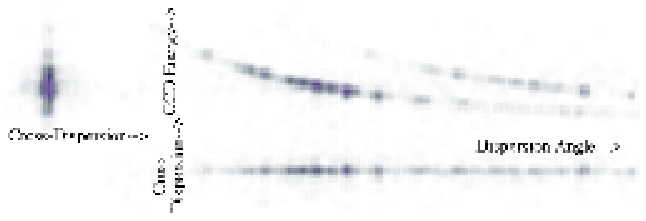}{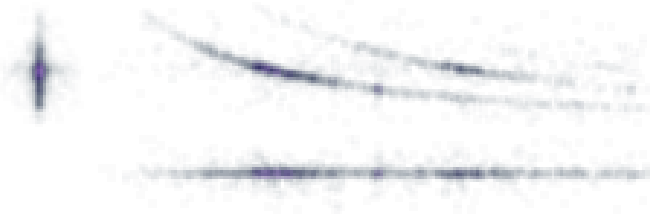}
\includegraphics[width=3in]{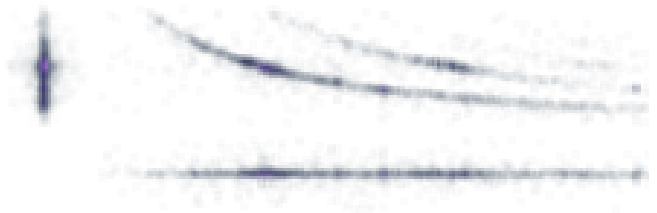}
\caption[Statistical simulation]
{\label{fig:statsim} (Three panels) Three simulations of a thermal plasmas at electron
  temperatures of 0.5, 1.0, and 1.5 keV emitted from an unresolved source using the RGS Monte Carlo.
  Each plot is segmented in the same way as in Figure 11 and 12.  The bright
  spots represent photons from strong emission lines.  The simulations are
  compared in Figure 14.  Each simulation has approximately 30,000 photons.
}
\end{figure}

\begin{figure}
\includegraphics[width=3in]{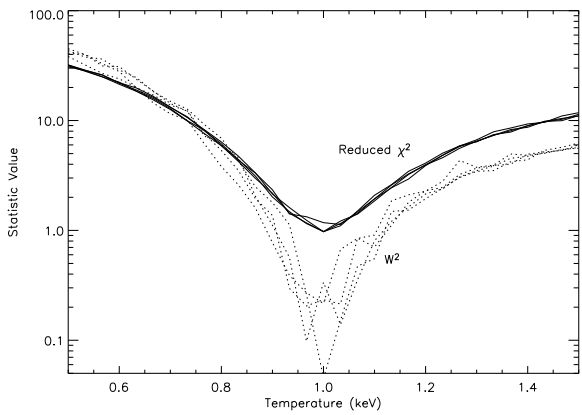}
\caption[Statistic comparison]
{\label{fig:statplot} 
The two-sample $\chi^2$ and the Cram\'{e}r-von Mises statistic, $W^2$, (see text)
calculated on the one-dimensional extracted spectrum by comparing the 1 keV
simulation in Figure 13 with various other simulations at other temperatures.
The statistics correctly achieve the minimum at 1 keV and fluctuate around
their 50th percentile values (1.0 for $\chi^2$ and 0.12 for $W^2$).  The
fluctuations can be diminished if more events are simulated.}
\end{figure}

\end{document}